\documentstyle[12pt,epsfig,multirow]{article}

\newcommand{\be}{\begin{eqnarray}}
\newcommand{\ee}{\end{eqnarray}}
\newcommand{\etal}{{\it et al. }}
\def\nue{{\nu_e}}
\def\anue{{\bar\nu_e}}
\def\numu{{\nu_{\mu}}}
\def\anumu{{\bar\nu_{\mu}}}

\newcommand{\chr}{\mbox{$\breve{\rm C}$erenkov~}}

\def\gsim{\:\raisebox{-0.5ex}{$\stackrel{\textstyle>}{\sim}$}\:}

\newcommand{\sss}{\sin^2 \theta_{12}}

\def\ltap{\ \raisebox{-.4ex}{\rlap{$\sim$}} \raisebox{.4ex}{$<$}\ }

\begin{document}

\thispagestyle{empty}
\begin{flushright}
\texttt{HRI-P-08-05-001}\\
\texttt{SINP/TNP/2008/11}\\
\texttt{TIFR/TH/08-20}
\end{flushright}
\bigskip

\begin{center}
{\Large \bf Effect of Collective Flavor Oscillations on the \\Diffuse Supernova 
Neutrino Background} 

\vspace{.5in}

{\bf {Sovan Chakraborty$^{\star, a}$, 
Sandhya Choubey$^{\dagger, b}$, Basudeb Dasgupta$^{\sharp, c}$,
Kamales Kar$^{\star, d}$}} 
\vskip .5cm
$^\star${\normalsize \it Saha Institute of Nuclear Physics,}\\
{\normalsize \it 1/AF Bidhannagar, Kolkata 700064, India}\\
\vskip 0.4cm
$^\dagger${\normalsize \it Harish-Chandra Research Institute,} \\
{\normalsize \it Chhatnag Road, Jhunsi, Allahabad  211019, India}\\
\vskip 0.4cm
$^\sharp${\normalsize \it Tata Institute of Fundamental Research,}\\
{\normalsize \it Homi Bhabha Road, Mumbai 400005, India} 
\vskip 1cm

{\bf ABSTRACT}
\end{center}

Collective flavor oscillations driven by neutrino-neutrino 
interactions inside core-collapse supernovae have now been 
shown to drastically alter the resultant 
neutrino fluxes. This would in turn significantly affect the 
diffuse supernova neutrino background (DSNB), created by 
all core-collapse supernovae that have exploded in the past. 
In view of these collective effects, we re-analyze the potential 
of detecting the DSNB in currently running and planned 
large-scale detectors meant for detecting both $\bar\nu_e$ 
and $\nu_e$. We find that the event rate can be different from previous estimates by 
upto $50\%$, depending on the value of $\theta_{13}$. The next generation detectors should 
be able to observe DSNB fluxes. Under certain conducive conditions, 
one could learn about neutrino parameters. For instance, it might 
be possible to determine the neutrino mass hierarchy, even if 
$\theta_{13}\rightarrow 0$.

\noindent

\vfill

\noindent $^a$ email: sovan.chakraborty@saha.ac.in

\noindent $^b$ email: sandhya@hri.res.in

\noindent $^c$ email: basudeb@theory.tifr.res.in

\noindent $^d$ email: kamales.kar@saha.ac.in

\newpage

\section{Introduction}
Observation of neutrino signal from a core-collapse supernova (SN) is
expected to contribute significantly towards determination of the neutrino
mass hierarchy and the mixing angle $\theta_{13}$.  
It is also expected to shed light on SN shockwave dynamics and set stringent
bounds on the existence of sterile neutrinos. Thus a lot of attention has been
devoted to this scenario of resonant neutrino conversions in a SN 
\cite{Mikheev:1986if,fuller-mayle-wilson-schramm-ApJ322}. 
The work has focussed on the determination of mass hierarchy and 
signatures of a non-zero $\theta_{13}$ 
\cite{ds,lunardini-smirnov-0302033}, 
Earth matter effects on the neutrino fluxes when they
pass through matter \cite{dighe-keil-raffelt-jcap0306005,
dighe-keil-raffelt-jcap0306006,dighe-kachelriess-raffelt-tomas-jcap0401},
shock wave effects on observable neutrino spectra and their
model independent signatures \cite{fuller}-\cite{huber}. 
Recently, possible interference effects for multiple resonances 
\cite{dasgupta-dighe-0510219}, sterile neutrinos 
\cite{choubey-harries-ross-0605255}, the role of turbulence in 
washing out shock wave effects 
\cite{fogli-lisi-mirizzi-montanino-0603033, 
friedland-gruzinov-0607244, Choubey:2007ga}, 
and time variation of the signal \cite{Kneller07} have 
also been explored. Observational data on SN neutrinos 
come only from the SN1987A \cite{Bionta87, Alexeyev87, Hirata87}, 
and while it confirms broad expectations, the small number of 
events limits our ability to draw detailed inferences.

However, supernovae are relatively rare in our galaxy with an estimated rate of about
$1-3$ per century \cite{Diehl:2006cf}, which prompts consideration of the
alternative strategy to detect neutrinos from supernovae that are further
away. Neutrinos accumulated in the Universe from all the SN explosions 
in the past and present epoch form a cosmic background, known as the 
diffuse supernova neutrino background (DSNB) or supernova relic neutrinos \cite{Bisno:84,Krauss:1983zn,Woosley:86}. The expected flux of these DSNB 
neutrinos depends mainly on the SN rate and the``typical" flavor 
dependent flux of neutrinos from supernovae. 

The SN rate can be either determined directly \cite{Mannucci:2007pb} or from
the cosmic star formation rate, which is measured using a variety of
ways like galaxy luminosity function of rest-frame ultraviolet radiation
\cite{Lilly96}--\cite{Steidel99}, far-infrared/sub-millimeter dust emission
\cite{Hughes98,Flores99} and near-infrared H$\alpha$ fluorescent line emission
\cite{Gallego95}--\cite{Glazebrook99} and radio emission
~\cite{Archibald:2000rw}. Though these techniques suffer from various ambiguities
and complications like dust extinction \cite{Somerville01}, careful studies
have enabled a precise determination of the star formation rate \cite {Strigari:2005hu}.

The typical neutrino flux from a SN on the other hand is not experimentally
available. The data from SN1987A do not allow a clean determination of the
spectral parameters \cite{Mirizzi:2005tg}. So one has to resort to using
primary neutrino fluxes predicted using SN simulations. Fluxes predicted by
different groups (and sometimes different simulations by the same group) are
at considerable variance, because of their different physics input
\cite{garching, livermore}. Additionally, these primary fluxes are further
mixed by flavor conversions as they stream through the SN, thus requiring
knowledge of not only the typical primary spectra, but also the typical SN
density profiles. The emitted neutrino fluxes are therefore ridden with
uncertainities at present. However, they could be made precise with more
sophisticated simulations or observation of a galactic SN.

Estimation of the DSNB flux has been performed in previous literature, with
varying approaches and results \cite{Strigari:2005hu},
\cite{Malaney:1996ar}--\cite{Volpe:2007qx}. Most studies have focussed on DSNB
detection via $\anue$ scattering off protons at water \chr detectors
\cite{Ando:2004sb} and large liquid scintillator detectors
\cite{Wurm:2007cy}. On the other hand, 
$\nue$ detection has been considered at
a liquid argon detector \cite{Cocco:2004ac} and 
at Sudbury Neutrino Observatory (SNO) 
\cite{Beacom:2005it, Lunardini:2006sn}. 
In \cite{Volpe:2007qx}, authors have performed a detailed 
comparative study of $\nue$ detection in 
different future large scale observatories -- by 
interaction of $\nue$ on oxygen in water \chr detectors, 
on carbon in liquid scintillator detectors and on argon in 
liquid argon detectors. 
Experimentally, the best upper
limits at 90 $\%$ C.L. of 6.8 $\times$ 10$^3$ $\nu_e$ 
cm$^{-2}$s$^{-1}$  (25~MeV$< E_{\nu_e}<50$ MeV) and 1.2 $\bar{\nu}_{e}$
cm$^{-2}$s$^{-1}$ ($E_{\bar{\nu}_e}>19.3$~MeV) come from the Liquid Scintillation Detector (LSD) \cite{Aglietta:1992yk} and
the Super-Kamiokande (SK) detectors
\cite{Malek:2002ns} respectively. 
However, stronger bounds can be placed on these fluxes, albeit using somewhat 
indirect arguments \cite{Lunardini:2006sn, Lunardini:2008xd}.
Some of the theoretical estimates of the DSNB fluxes
predict event-rates for $\anue$ that are tantalizingly close to detection, 
{\it e.g.},
the observational upper limit set by the SK collaboration
\cite {Malek:2002ns}. The prospects for discovery thus seem promising if a
large water \chr detector like SK is loaded with $0.02\%$ ${\rm
GdCl}_{3}$ \cite{Beacom:2003nk} or  if one or more of the proposed next
generation detectors become available.

The study presented in this work removes a common assumption, made for all
previous estimates, that neutrino-neutrino interactions are too feeble to
cause any flavor conversion. Although these interactions had been studied in
previous literature \cite{Pantaleone:1992eq}-\cite{Pastor:2001iu}, it has only
very recently been appreciated that they induce sizable flavor conversion
in supernovae \cite{Duan:2005cp, Duan:2006an}. Motivated by this interesting result,
the effects of these neutrino-neutrino interactions have been explored in the
context of supernovae, in a series of papers
\cite{Hannestad:2006nj}-\cite{Dasgupta:2008my}. The interesting aspect of
these conversions is that the neutrinos and antineutrinos of different
energies undergo conversions together, and are almost in-phase. Therefore
these conversions are referred to as being ``collective". The effect of these
interactions becomes negligible beyond the first few hundred kilometers, when
the neutrino densities become much lower. The end of these collective effects
is marked by a complete (or step-wise) swapping between the flavor spectra of
antineutrinos (or neutrinos) for inverted hierarchy (IH) \cite{Raffelt:2007cb,
  Duan:2007fw, Raffelt:2007xt}. Further out from the centre of the star, the
traditional picture of flavor evolution by Mikheyev-Smirnov-Wolfenstein (MSW) 
conversion is not changed, except that the primary fluxes emitted at the 
neutrinosphere undergo the above-mentioned ``pre-processing" due to the 
collective effects. The fluxes emitted by a SN have already been calculated 
in a three-flavor framework, including
the effect of collective oscillations \cite {Dasgupta:2007ws}. In this paper,
we take these SN neutrino fluxes calculated in \cite {Dasgupta:2007ws}, 
the SN rate deduced from the cosmic star formation rate calculated by Beacom \etal \cite {Strigari:2005hu}, 
and the standard $\Lambda$-CDM cosmological model \cite{Dunkley:2008ie} as inputs 
to calculate the DSNB flux. The expected DSNB flux in the case of IH turns 
out to be quite different from those contained in previous works that disregarded 
collective effects. Thus the prospects of DSNB detection at antineutrino and/or neutrino
detectors are modified. We report the DSNB fluxes and their observability,
with and without neutron tagging, at present and proposed detectors. These
changed expectations have impact on limits that can be set on non-standard
models and interactions of neutrinos as well.

The paper is organised as follows. In Section 2, we outline the dependence on
the DSNB event-rate at Earth as a function of the cosmic star formation rate 
and the primary neutrino fluxes and mention our choice of
models for the same. We discuss our choice of detectors.  
In Section 3, we calculate the event-rates, dependent on
the above choices, and present the results. We conclude the paper in Section
4, by summarizing our results and giving an outlook on the impact of these
results.

\section{DSNB in Terrestrial Detectors}

From the early times to the present date, SN explosions
have been fairly common events in the Universe. These explosions have injected a 
large number of neutrinos, with energies of tens of MeV, in the Universe. These 
neutrinos have created a diffuse background of SN neutrinos known as 
diffuse supernova neutrino background. Evidently, the DSNB flux depends on two ingredients:
\begin{itemize} 
\item The rate of SN explosions $R_{SN}(z)$, as a function of cosmological redshift $z$.
\item The differential flux of neutrinos $F_{\nu}(E_{\nu})$, from a typical core-collapse event at redshift $z$.
\end{itemize}
The differential flux of neutrinos $F_{\nu}(E_{\nu})$ depends on the 
primary neutrino fluxes $F^0_{\nu}(E_{\nu})$, emitted from the 
neutrinosphere, which get modified due to 
\begin{itemize}
\item Collective effects, i.e. neutrino-neutrino self interaction, close to the neutrinosphere.
\item MSW effects, i.e matter driven neutrino oscillations in the SN mantle and envelope.
\end{itemize}

The total differential DSNB flux arriving at terrestrial detectors, 
expressed as the number of neutrinos of flavor $\nu$ (where
$\nu=\nu_e,\nu_\mu,\nu_\tau$ and antineutrinos are denoted with a bar overhead) 
arriving per unit area per unit time per unit energy, due to all supernovae 
in the Universe up to a maximum redshift $z_{max}$, is
\begin{eqnarray}
F'_{\nu}(E_{\nu}) 
=\int_{z_{max}}^0 (dz\,{\frac{dt}{dz}})\,(1+z)\,
R_{SN}(z)\,F_{\nu}((1+z)E_{\nu})~.
\end{eqnarray}
Here $E_{\nu}$ is the neutrino energy at Earth and $R_{SN}(z)$ is the SN rate
per comoving volume at redshift $z$. For our numerical calculations 
we have assumed $z_{max}=7$. 
Note that the factor $(1+z)$ in the
neutrino spectrum $F_{\nu}((1+z)E_{\nu})$ incorporates the redshift of the
energy spectrum.

From the Friedmann equation for a flat universe we have 
\begin{equation}
\frac{dz}{dt}=-H_0(1+z)(\Omega_m(1+z)^3+\Omega_\Lambda)^{1/2}
~.
\end{equation} 
Thus the differential number flux of DSNB is
\begin{eqnarray}
F'_{\nu}(E_{\nu})
=\frac{1}{H_0}\,{\int_0^{z_{max}}}\,R_{SN}(z)\,
F_{\nu}((1+z)E_{\nu})
\frac{dz}{\sqrt{(\Omega_m(1+z)^3+\Omega_\Lambda)}}
~.
\end{eqnarray}
For the standard $\Lambda$-CDM cosmology, we have
\begin{equation}
\Omega_m = 0.3~;~\Omega_\Lambda = 0.7~{\rm and}~H_0 = 70~h_{70}~{\rm km}~{\rm s}^{-1}~{\rm Mpc}^{-1}~.
\end{equation}
Therefore, we only need to know the SN rate $R_{SN}(z)$ and the differential flux of neutrinos $F_{\nu}(E_{\nu})$, from a typical core-collapse event to calculate the DSNB flux at Earth.

\subsection{The Cosmic Supernova Rate}

The SN rate $R_{SN}(z)$ is related to $R_{SF}(z)$, through 
the initial mass function $\varphi(m)$, which describes the
differential mass distribution of stars at formation 
\cite{Strigari:2005hu,Ando:2004hc}.
We assume that all stars that are more massive than $8M_\odot$ give rise to core-collapse events and die on a timescale much shorter than the Hubble time, and that the initial mass function $\varphi(m)$ is independent of redshift. This allows us to relate the star formation rate $R_{SF}(z)$ to the cosmic SN rate $R_{SN}(z)$ as
\begin{equation}
{R_{SN}(z)}=R_{SF}(z)
\frac{{\int_{8M_\odot}^{125M_\odot}\varphi(m)dm}}
{{\int_{0.1M_\odot}^{125M_\odot}\varphi(m)
m dm}}~.
\label{rsnz}
\end{equation}
For our estimates, we use the initial mass function from 
reference \cite{Baldry-Glazebrook}, i.e. 
\begin{eqnarray}
\varphi(m) \propto 
\Bigg\lbrace 
\begin{array}{lc}
m^{-1.50} & (0.1 M_{\odot} < m < 0.5 M_{\odot})\\ 
m^{-2.15} & (m > 0.5 M_{\odot})\\ 
\end{array}~.
\end{eqnarray}
Putting the above expression into Eq. (\ref{rsnz}) we find 
\begin{equation}
{R_{SN}(z)} = 0.0132~~R_{SF}(z) {M_{\odot}^{-1}}~.
\end{equation}
It should be noted that the factor connecting $ R_{SN}$ and $R_{SF }$ is 
quite insensitive to the upper limit of the integrations in Eq. (\ref{rsnz}). 

Recent careful studies on different indicators of the cosmic star formation rate 
have been used to calculate the $R_{SF}$ and its normalization. We use the 
cosmic star formation rate per comoving volume, $R_{SF}$, from the concordance 
model advocated in \cite{Hopkins-Beacom_06,Kistler_etal_07}, which is given by
\begin{equation}
{R_{SF}(z)}\propto\Bigg\{ \begin{array}{lc}{(1+z)^{3.44}} & z<0.97 \\
{(1+z)^{-0.26}} & 0.97<z<4.48\\
{(1+z)^{-7.8}} & 4.48<z\\
\end{array}\;,
\end{equation}
with the local star formation rate given by
\begin{equation} 
R_{SF}(0)=0.0197~{M_\odot}{{\rm yr}^{-1}}{{\rm Mpc}^{-3}}\;.
\end{equation}
This model satisfies the experimental upper limit on DSNB set by SK
\cite{Malek:2002ns}, and hence is known as the 
concordance model \cite{Strigari:2005hu, Bhattacharjee:1st}.

\subsection{Neutrino Fluxes from Core-collapse Supernovae}

\begin{figure}
\includegraphics[width=8.0cm,height=9.0cm,angle=270]{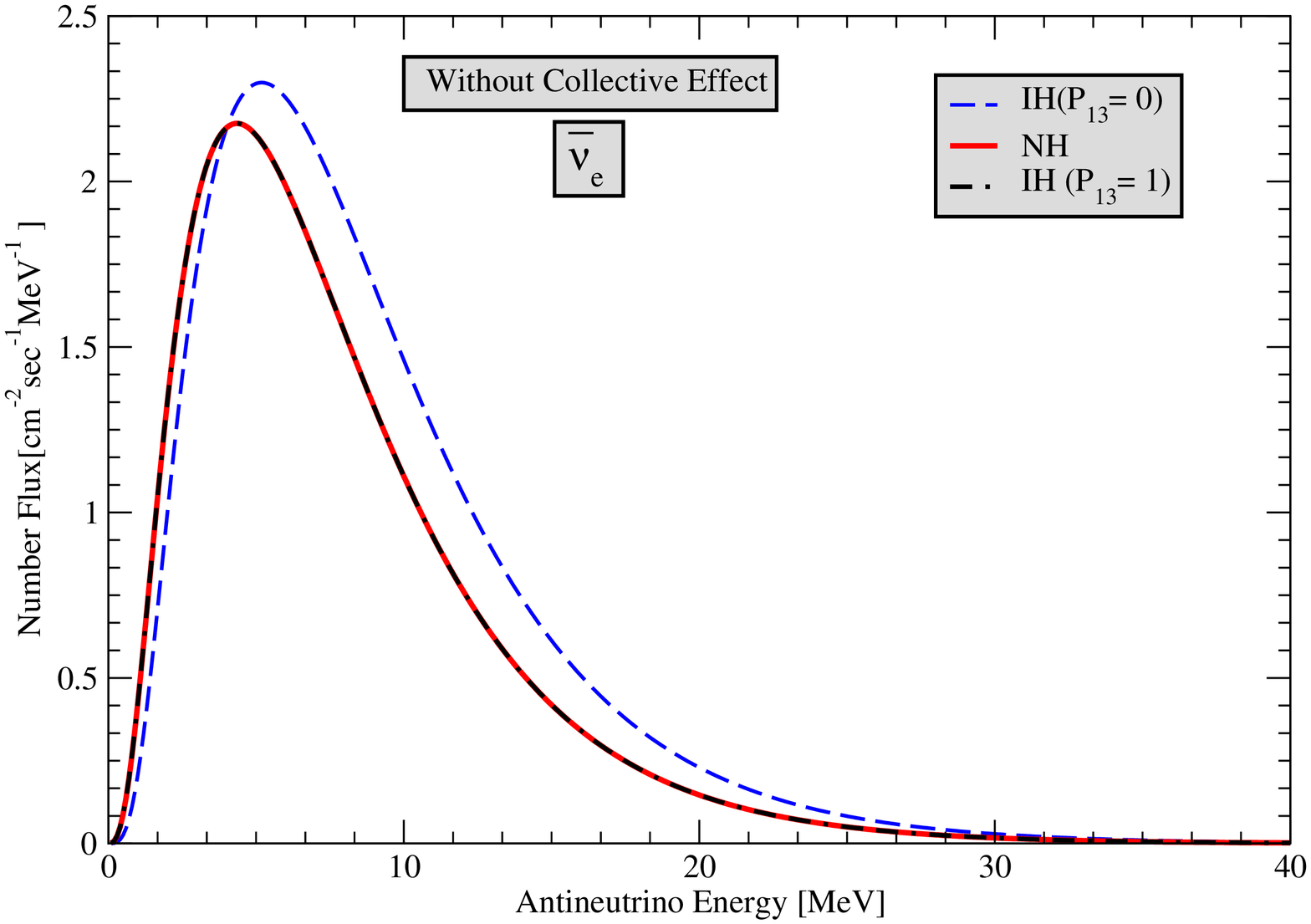}
\vglue -8.0cm \hglue 8.5cm
\includegraphics[width=8.0cm,height=9.0cm,angle=270]{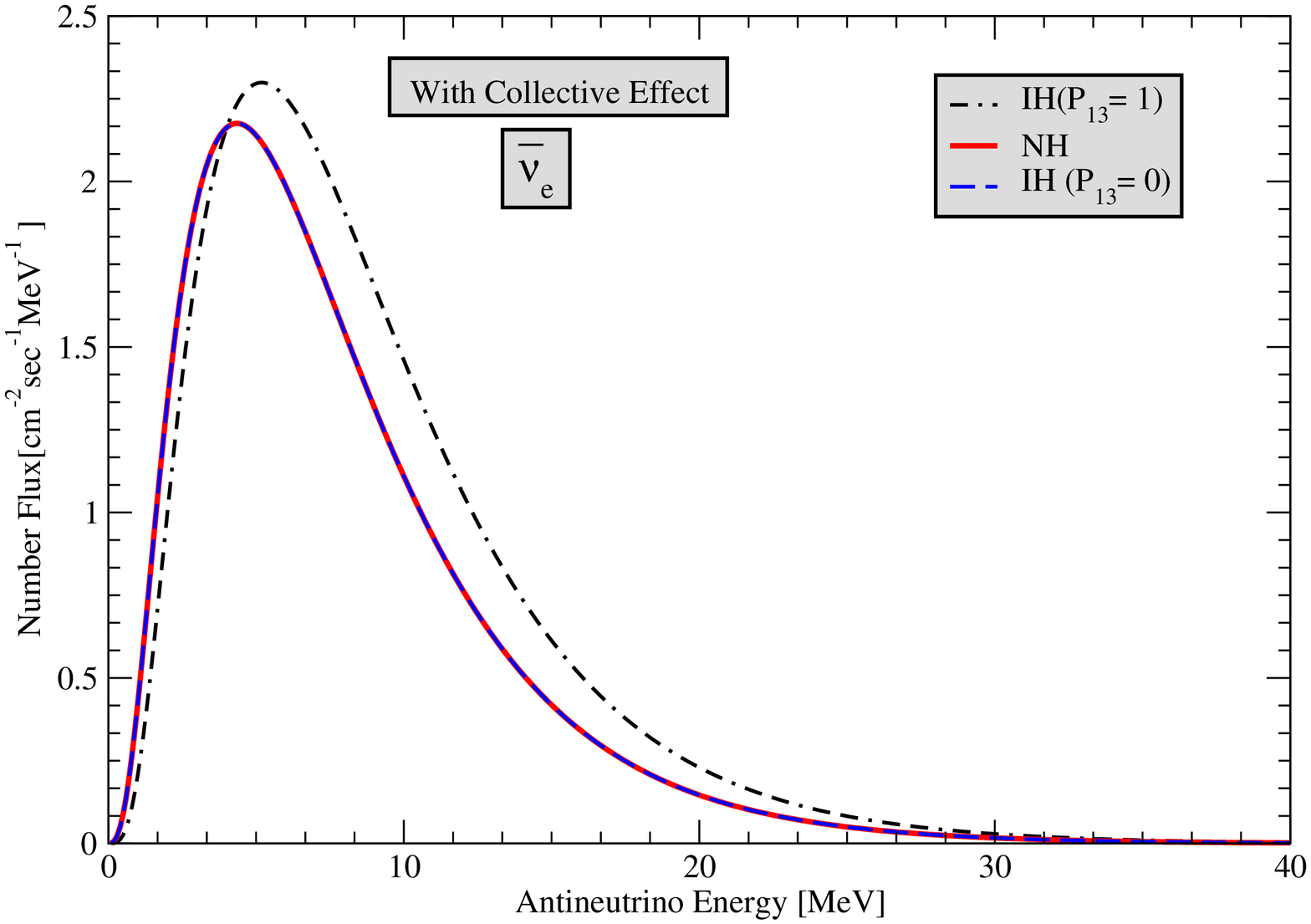}
\includegraphics[width=8.0cm,height=9.0cm,angle=270]{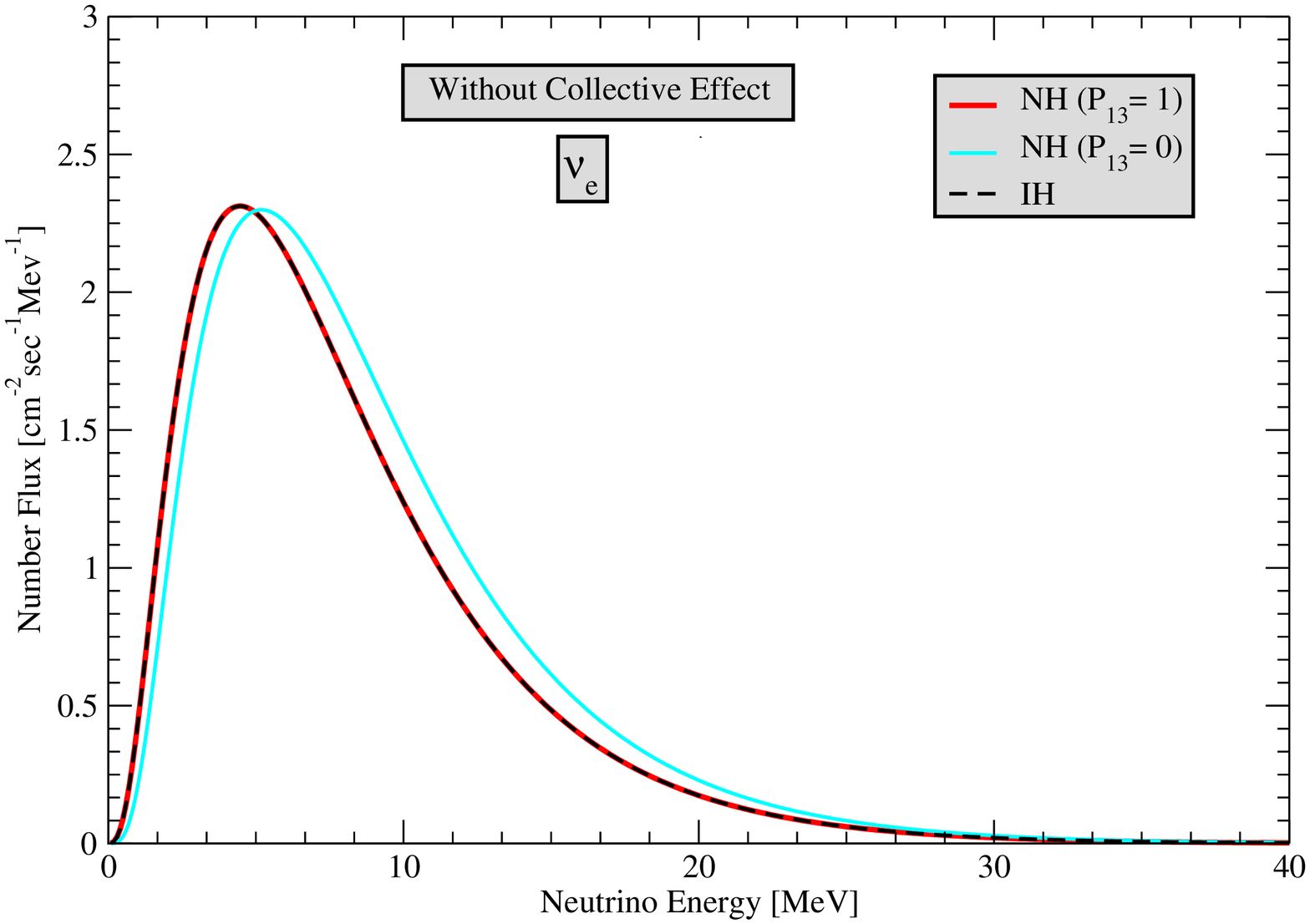}
\vglue -8.0cm \hglue 8.5cm
\includegraphics[width=8.0cm,height=9.0cm,angle=270]{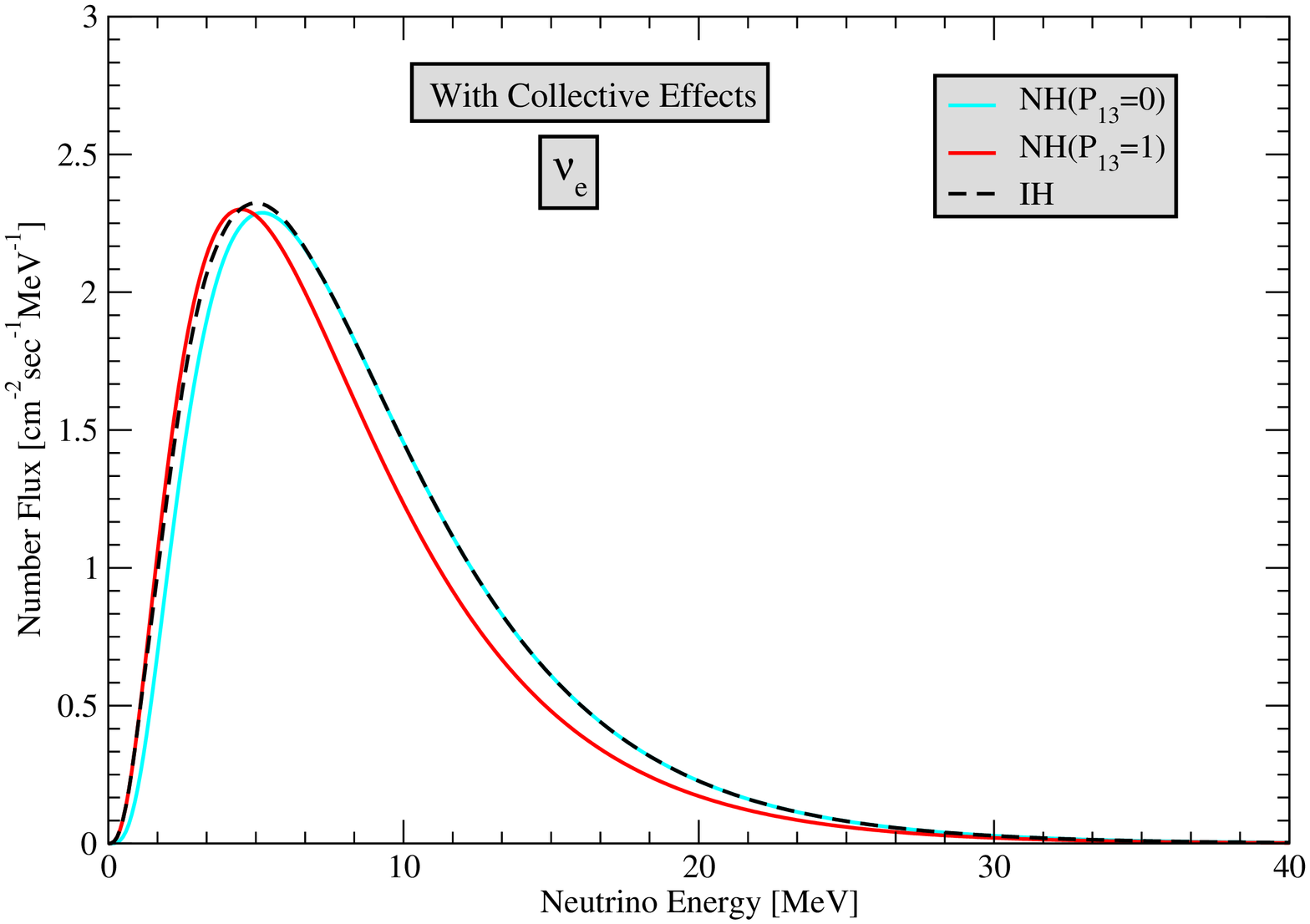}
\caption{\label{fig:flux}
The DSNB flux spectrum arriving at Earth as a function of the 
(anti)neutrino energy at Earth. The upper panels show the 
$\anue$ flux while the lower panels show the $\nue$ flux. 
The left panels correspond to the hypothetical case where 
we have only MSW matter effects in the SN while 
the right panels correspond to the case where we have both collective as 
well as MSW-driven flavor transitions. We have assumed 
that the initial neutrinos are given by the G1 model. 
}
\end{figure}

\subsubsection{Primary Neutrino Fluxes}
A core-collapse SN is typified by conversion of its gravitational
binding energy of about $10^{53}$ ergs to neutrinos and antineutrinos with
energies of tens of MeV. This translates to a total flux of more than
$10^{57}$ per SN explosion. The initial luminosity is about $10^{53}$ ergs/sec
(emitted purely as $\nu_e$ in the neutronization burst lasting for about $25$
msecs), and slowly reduces to about $10^{51}$ ergs/sec over the following
$\sim 10$ secs. Since the event-rate for DSNB is not very sensitive to the
neutronization burst, we can ignore it for our analysis \footnote{The
  neutrinos in the burst phase undergo complicated flavor conversions,
  particularly for a class of low-density SN with degenerate O-Mg-Ne cores
  \cite{Duan:2007sh, Lunardini:2007vn, Duan:2008za, Dasgupta:2008cd}. However, these issues are
  not likely to be important for DSNB due to the relative smallness of the
  integrated flux from the burst-phase.}. Subsequently, after the
neutronization burst, neutrinos and antineutrinos of all three flavors are
emitted with a pinched thermal spectrum, that is conveniently parametrized as
\cite{Keil:2002in}
\be
F_{\nu}^{0}(E_{\nu}) =
\frac{L_{\nu}^{0}}{\langle E_{\nu}
  \rangle^2}\frac{{\left(1+\zeta_{\nu}\right)}^{1+\zeta_{\nu}}}
{\Gamma(1+\zeta_{\nu})}\left({\frac{E_{\nu}}{\langle E_{\nu}
      \rangle}}\right)^{\zeta_{\nu}}\exp\left(-{(1+\zeta_{\nu})\frac{E_{\nu}}
{\langle E_{\nu} \rangle}}\right)~,
\ee
where $L_{\nu}^{0}$ is the luminosity in the flavor $\nu$, $\langle E_{\nu} \rangle$ is the average
energy of $\nu$, and $\zeta_{\nu}$ is the pinching
parameter at the neutrinosphere. As a notational convenience, since there is no
difference expected between the  $\mu$ and $\tau$ flavors for SN neutrinos, we
will refer to them together as $x$.

For our study, we take 3 sets of representative values of $\Phi_{\nu}^{0}=L_{\nu}^{0}/\langle E_{\nu} \rangle$, $\langle E_{\nu} \rangle$ and $\zeta_{\nu}$ motivated by SN simulations.
One simulation by the Lawrence Livermore group (LL), and two different simulations by the
Garching group (G1, G2) have been chosen for our estimates, as shown in Table
\ref{SNmodeltable}.

\begin{table}[ht]
\begin{center}
\begin{tabular}{llllll}
\hline
Model & $\langle E_{\nue} \rangle$ & $\langle E_{\anue} \rangle$&
$\langle E_{\nu_x, {\bar{\nu}_x}} \rangle$ & {\large 
$\frac{\Phi^0_{\nue}}{\Phi^0_{\nu_x}}$} &
{\large $\frac{\Phi^0_{\anue}}{\Phi^0_{\bar{\nu}_x}}$}\\
\hline
LL & 12 & 15 & 24 & 2.0 & 1.6 \\
G1 & 12 & 15 & 18 & 0.8 & 0.8 \\
G2 & 12 & 15 & 15 & 0.5 & 0.5 \\
\hline
\end{tabular}
\caption{The parameters of the used primary neutrino spectra models
  motivated from SN simulations of the Garching (G1, G2) and the
  Lawrence Livermore (LL) group. We assume 
$\zeta_{\bar{\nu}_x}=4$ and $\zeta_\anue=3$. The total luminosity is 
chosen to be $3\times10^{53}$ erg.} 
\label{SNmodeltable}
\end{center}
\end{table}

Note that the LL simulation obtained a large hierarchy $\langle
E_{\nue}\rangle<\langle {E}_{\anue}\rangle<\langle E_{\nu_x}\rangle\approx
\langle {E}_{\bar{\nu}_x}\rangle$, and an almost complete equipartition of
energies between flavors. The Garching simulations predict a smaller hierarchy
between the average energies, incomplete equipartition, and increased spectral
pinching. The differences in the values of these parameters arise from the
different physics inputs.

\subsubsection{Collective Effects and MSW Transitions}
The primary fluxes (at the neutrinosphere) are further processed by collective
effects and MSW conversions before they get emitted from the SN\footnote{The detailed 
picture of collective effects presented herein is valid only for initial spectra that
resemble the LL model. However we are interested in seeing the maximum effect that these 
new effects can cause, and for that purpose it suffices to ignore more complicated features 
in the spectrum \cite{Dasgupta:2009mg} for G1 or G2 like models.}.

Near the neutrinosphere, due to the large neutrino density, the
neutrino-neutrino interaction energy is very large. This ensures that the
neutrinos exhibit synchronized oscillations, i.e. neutrinos of all energies
oscillate coherently with the average frequency. These oscillations do not
give rise to any effective flavor conversion since the effective mixing angle
is highly supressed due to the large MSW potential. As the neutrinos stream
outward, the neutrino density becomes smaller, and bipolar oscillations begin
to take place. In the case of IH, these oscillations have large amplitude even
for a vanishingly small mixing angle. These oscillations thus can lead to a
complete swapping of the $\anue$ and $\bar{\nu}_x$ spectra. The $\nue$ and
$\nu_x$ spectra cannot swap completely, because of lepton number conservation,
and the swap occurs only above a certain energy $E_c$, giving rise to a spectral
split \cite{Raffelt:2007cb}. Eventually, beyond a few hundred kilometers, the neutrino-neutrino
interaction energy becomes negligible, and collective effects cease to be
important.

Thus for normal hierarchy (NH), the collective effects do not affect the
fluxes significantly and only MSW conversions are at work. In particular, the
MSW resonances affect the $\nu_e$ flux, while the $\anue$ flux remains almost
unaffected. 
For IH, the collective effects swap the $\nu_e$ and $\nu_x$ above a certain
energy $E_c$, determined by lepton number conservation \cite{Raffelt:2007cb,Duan:2008za, Dasgupta:2008cd}. Assuming
$F^{0}_{\nu_x}={F}^{0}_{\bar{\nu}_x}$, the split-energy $E_c$ is given in the
adiabatic approximation by the implicit equation
\be
\int_{0}^{E_c}dE \; \left(F^{0}_{\nue}(E)-{F}^{0}_{\nu_x}(E)\right)=\int_{0}^{\infty}dE \; \left(F^{0}_{\nue}(E)-{F}^{0}_{\anue}(E)\right)~.
\ee
On the other hand for antineutrinos, all $\anue$ and $\bar{\nu}_x$ are
swapped. This pre-processed flux now undergoes the traditional MSW conversions
which now affect the $\anue$ flux, and not the $\nue$ flux. The neutrinos then
travel independently (while getting redshifted) as mass-eigenstates until
they reach Earth, wherein they are detected as flavor eigenstates before or
after having undergone regeneration inside the Earth. The fluxes of $\nue$ and
$\anue$ arriving at Earth are given in Table \ref{fluxtable}, as calculated in
\cite{Dasgupta:2007ws}. The quantities such as 
$F_{\nu_{\alpha}}^0$, are the initial SN neutrino fluxes while 
$F_{\nu_{\alpha}}$ are the resultant fluxes emerging from the SN at redshift $z$. 
\begin{table}[ht]
\begin{tabular}{lcl}
\hline
Normal hierarchy & $\phantom{space}$ & Inverted hierarchy \\ 
\hline
$F_{\nue}= s_{12}^2\left(P_{13} F^{0}_{\nue}+(1-P_{13}) 
F^{0}_{\nu_x}\right) + c_{12}^2 F^{0}_{\nu_x}$  & &
$F_{\nue}= \Bigg\{ \begin{array}{lc}
s_{12}^2 F^{0}_{\nue}+c_{12}^2F^{0}_{\nu_x} & (E<E_c) \\
F^{0}_{\nu_x} & (E>E_c)\\ 
\end{array} $ \\
\vspace{0.1cm}
$F_{\anue}= c_{12}^2F^{0}_{\anue} + s_{12}^2F^{0}_{\bar{\nu}_x}$ &  & 
$F_{\anue}= s_{12}^2F^{0}_{\bar{\nu}_x} + c_{12}^2 \left((1-P_{13})
F^{0}_{\anue} + P_{13} F^{0}_{\bar{\nu}_x}\right)$\\
\hline
\end{tabular}
\caption{Electron neutrino and antineutrino spectra
emerging from a SN. See \cite{Dasgupta:2007ws} for a prescription for calculating these final spectra in terms of the primary spectra.}
\label{fluxtable}
\end{table}
The quantities $s_{12}^2$ and $c_{12}^2$ stand for 
$\sss$ (taken to be $0.3$ for numerical studies) and $\cos^2\theta_{12}$ respectively and 
$P_{13}$ is the effective jump probability between the neutrino
mass eigenstates due to the MSW resonance(s), and takes a value between $0$ and
$1$. The value of $P_{13}$ is approximately $0$ if $\theta_{13}$ is large
(i.e. $\theta_{13}\gsim 6$ degrees) and for smaller values of $\theta_{13}$ it
has a non-trivial dependence on energy and time, due to multiple resonances
\cite{dasgupta-dighe-0510219, Kneller07} and turbulence
\cite{fogli-lisi-mirizzi-montanino-0603033, friedland-gruzinov-0607244,
  Choubey:2007ga}. However, due to the small number of events, we can probably
neglect these sub-leading effects that occur in the small time-window when the
shockwave is in the resonance region. 

To calculate the DSNB flux at Earth $F'(E_\nu)$, we need to integrate the fluxes in Table \ref{fluxtable}, correctly redshifted and weighted by the SN rate $R_{SN}(z)$, over 
redshift $z$. We show in Fig. \ref{fig:flux} the DSNB 
$\anue$ (upper panels) and $\nue$ (lower panels) fluxes 
arriving on Earth as a function of their (anti)neutrino energy
at Earth. We have assumed the G1 model for generating this figure. 
Note that the energy spectrum gets degarded to 
smaller energies due to redshift. The left panels show the 
predicted fluxes when one takes both collective as well as 
MSW transitions into account. To bring out the contrast with what the 
situation was earlier, we show in the right panels the 
predicted fluxes if one does not take collective effects. 
We can see that for NH the prediction have remained 
the same even after collective effects were taken, 
whereas for IH the fluxes are completely different.


\subsection{Terrestrial Detectors}

An array of existing and planned detectors could 
catch the DSNB neutrinos. In what follows, we will 
consider in particular three types of detectors for 
observing DSNB $\anue$:
\begin{itemize}
\item Water \chr detectors 
\item Liquid scintillator detectors
\item Gadolinium loaded water \chr detectors
\end{itemize}
Detection of $\nue$ is more difficult. Both 
water and liquid scintillator detectors can in principle 
detect $\nue$ (as well muon and tau flavored neutrinos and 
antineutrinos). In water \chr detectors this can be done through 
neutrino-electron scattering. On the other hand, in liquid scintillators   
in addition to the neutrino-electron scattering, 
one can detect $\nue$ through charged current interaction 
on $^{12}C$, while the other species can be detected through 
the neutral current interaction 
on $^{12}C$. However, the cross-section for these processes 
are rather low. Another detector technology that has been proposed 
for detecting $\nue$ is to use a high $Z$ material, such as 
lead (and/or iron), interleaved with scintillators. 
Among such proposals are the OMNIS/ADONIS projects 
and the HALO experiment at SNOLAB \cite{snolab}.
Therefore the only chance for detecting the 
$\nue$ DSNB would be in a reasonably large 
\begin{itemize}
\item Liquid argon detector 
\end{itemize}

\subsubsection{Water \chr Detectors}

An upper bound on the DSNB flux already exists from non-observation of these 
neutrinos at the SK experiment \cite{Malek:2002ns}. Using $1496$ days of data
with $22.5$ kton of fiducial volume, the DSNB flux has been constrained to be
less than $1.2$ cm$^{-2}$ s$^{-1}$ for 
$19.3 {~\rm MeV} < E_\nu < 30 $ MeV. 
SK is still running and could provide further 
constraint or evidence for DSNB fluxes in the future. 
Megaton water detectors with fiducial 
volume in the ballpark of $500$ kton have been planned 
in Japan (Hyper-Kamiokande (HK)) \cite{hyperK}, 
Europe (MEMPHYS) 
\cite{memphys}, and USA (UNO) \cite{uno}. These have been 
proposed to serve as the far detector for long baseline 
experiments with powerful accelerator beams. 
At the same time, they would be used to study neutrinos 
from natural sources, such as the Sun, atmosphere and 
nearby supernovae. In particular, they will be useful tools for 
the observation of DSNB fluxes. While in principle
water detectors can detect neutrinos and antineutrinos
of all flavors, the easiest to observe is $\anue$, 
which is captured on protons via the inverse 
beta decay process
\be
\anue + p \rightarrow e^+ + n
~.
\label{eq:anuep}
\ee
The emitted positron is observed through the 
\chr cone produced by it. 
The ``true'' positron energy is approximately related to the neutrino 
energy by $E_\nu - 1.3$ MeV.
The other types of neutrino species would scatter 
electrons and thereby could also be detected. However, 
the cross-section for neutrino-electron scattering is 
much lower 
compared to the reaction (\ref{eq:anuep}). Therefore, 
in this paper we will consider the detection of 
only $\anue$ in water \chr detectors. The number 
of events per kton of detector mass is given as 
\be
N_e = n_T\,T\,\int_{0}^{\infty} ~ dE_\nu
\int_{E_e^{low}}^{E_e^{up}} ~ dE_e
\,\,\,F'_{\nu}(E_\nu) \sigma(E_\nu) R(E_\nu, E_e)
~,
\label{eq:events}
\ee
where $n_T$ is the number of protons in a kton of 
detector mass, $T$ is the total 
exposure time, $E_e$ the measured positron energy,
$ E_e^{low}$ is the lower energy threshold, 
$E_e^{up}$ is the upper energy threshold, 
$F'_{\nu}(E_\nu)$ is the DSNB flux at Earth, 
$\sigma(E_\nu)$ is the cross-section and 
$R(E_\nu-1.3, E_e)$ is the energy resolution of the 
detector. For the energy resolution we assume a Gaussian form 
\be
R(E_\nu,E_e) = \frac{1}{\sqrt{2\pi}\sigma_E}
\exp \bigg(\frac{-(E_\nu-1.3 - E_e)^2}{2\sigma_E^2}\bigg)
~,
\label{eq:resol}
\ee
where all quantities are given in units of MeV 
and  $\sigma_E$ is the half width at half 
maximum (HWHM). For the water \chr detector we use 
\be
\sigma_E = 0.47\sqrt{(E_\nu({\rm MeV}) - 1.3)}
~.
\label{eq:eresolWC}
\ee

From Fig. \ref{fig:flux} we can see that the DSNB fluxes 
being redshifted,  
arrive on Earth predominantly within the energy window 
$E_\nu = (0 - 35)$ MeV, above which the fluxes are negligible.  
In this energy range water \chr detectors also register 
events coming from a myriad of other sources. The main 
sources of particles which would imitate the 
DSNB signal include reactor $\anue$,  
atmospheric $\nue$ and $\anue$, 
solar $\nue$, spallation products induced 
by cosmic ray muons, and neutrinos from ``invisible muons'' 
produced by atmospheric $\numu$ and $\anumu$. These 
form a background for the DSNB signal. 
Events due to reactor $\anue$ appear roughly in the energy 
range $(1.8-8)$ MeV and these events can be estimated using 
the information from reactor power and their distances
from the detector. In the case of SK for instance, it will be 
even easier to estimate them since KamLAND \cite{kl2008} directly 
observes these events.  
The events due to atmospheric $\nue$ and $\anue$ are 
expected to be lower compared to those due to DSNB 
below $E\simeq 30 $ MeV. 
Number of events expected from atmospheric $\nue$ and $\anue$
can be anyway estimated using the predicted fluxes at these 
energies and can be included in the analysis of DSNB events. Therefore, 
these events do not pose a very serious threat to the DSNB analysis. 
Events due to neutrinos coming from the Sun fall in the 
energy range $E_\nu \ltap 20$ MeV and 
can also be estimated fairly well 
using the fluxes from the standard solar model as well as 
from the direct observation of the $^8$B fluxes at SNO \cite{Aharmim:2005gt}. 
These neutrinos can also be 
identified in the detector from their directionality. 
Indeed these are the solar neutrino events that SK observes. 
Therefore, these events do not pose a serious threat to DSNB 
observation either. The type of events which 
cause a serious concern are the ones produced from spallation. 
These events are typically important in the energy window 
relevant for solar neutrinos, {\it viz.} for $E \ltap 20$ MeV. 
The SK collaboration in their paper \cite{Malek:2002ns} show that
after suitable cuts there are almost no spallation events 
above $E_e > 18$ MeV. The lower threshold for the 
neutrino energy is hence restricted to $E_\nu \geq 19.3$ MeV. 
The upper limit is taken as 30 MeV. 

Despite the different cuts and selection criteria there are 
two sources of neutrinos which still appear as backgrounds 
for the DSNB detection. The first has already been discussed above -- 
the $\nue$ and $\anue$ events from atmospheric neutrinos. These 
background events have to be estimated using the detector Monte 
Carlo. 
The second type of background comes from  ``invisible muons'' 
produced by atmospheric $\numu$ and $\anumu$. These are events 
where atmospheric $\numu$ and/or $\anumu$ produce muons with 
kinetic energy less than $53$ MeV, which is the threshold for 
emitting \chr photons. These muons/antimuons 
therefore pass undetected and 
eventually decay into electrons/positrons which are observed by 
the detector. Estimates for the background due to both these 
sources have been made by the SK collaboration and can be 
found in \cite{Malek:2002ns}.

\subsubsection{Liquid Scintillator Detectors}

Number of events expected in liquid scintillator are also 
given by Eq. (\ref{eq:events}). 
The predominant reaction is $\anue$ capture of protons 
(cf. reaction (\ref{eq:anuep})). The other detection reactions 
in liquid scintillators are charged and neutral current 
scattering off electrons, 
charged current capture of 
$\nue$ and $\anue$ on $^{12}$C, neutral current 
break-up of $^{12}$C 
(see \cite{klcs12c} for reactions of $^{12}$C)\footnote{Liquid 
scintillator detectors can also detect 
the DSNB $\nue$ flux by their charged 
current interactions on $^{12}$C 
\cite{Volpe:2007qx}.}
and neutral current scattering off 
protons \cite{beacomKLepscatt}. However, the cross-section 
for these processes is small, especially at low energies 
\cite{Volpe:2007qx,borexinosn,snklus}, and we reiterate that due to redshift 
the DSNB fluxes are peaked at lower energies. Therefore, 
even for liquid scintillators the main detection weapon is 
the reaction (\ref{eq:anuep}). However, compared to the water detectors,
liquid scintillators can use the reaction (\ref{eq:anuep}) 
more efficiently, whereby they tag the released 
neutron. While the positron is detected promptly, 
the neutron is captured by a proton in the detector, 
releasing a $2.2$ MeV photon which is detected 
in delayed coincidence after $180$ $\mu$s. 
This results in lesser problems with backgrounds, and 
liquid scintillator detectors can be used to observe the 
DSNB neutrino in the broader energy window of 
$E_\nu = (10 - 25)$ MeV \cite{Wurm:2007cy}. 

The other major difference between the liquid scintillator detector
and water \chr detector is in the energy resolution, which is 
much better for the former. The HWHM for liquid scintillator detectors 
is expected to be better than 
\be
\sigma_E = 0.1\sqrt{E_\nu{\rm (MeV)}-0.8}
~.
\ee

The KamLAND detector in Japan \cite{kl2008} 
and Borexino in Italy \cite{borexino} are 
the currently running liquid scintillator detectors. 
While KamLAND has a total mass of 1 kton, Borexino is much 
smaller and comprises of about $300$ ton of 
liquid scintillator. The detectors for the upcoming second 
generation reactor experiments designed to probe $\theta_{13}$ 
would be far too small to contribute to the study of DSNB neutrinos. 
However, one could look forward to proposals such as 
LENA \cite{Wurm:2007cy} which would be situated in the 
Pyhasalmi mine in Finland and 
is expected to have $50$ kton of liquid scintillator. Such a 
big liquid scintillator detector could collect sizable 
number of DSNB events and prove to be a pivotal player in 
this game. Another large liquid scintillator detector 
proposal is the Hanohano project in Hawaii \cite{hanohano}.

\subsubsection{Gadolinium Loaded Water \chr Detectors}

The neutron released in the reaction (\ref{eq:anuep}) 
when captured on protons emits only a $2.2$ MeV 
photon. This is below the detection threshold of water 
detectors and hence they cannot normally tag the 
released neutron by delayed coincidence, as liquid 
scintillators can. However, things could change 
dramatically if ${\rm GdCl}_3$ is dissolved into the water. Gadolinium 
has a large cross-section for neutron capture and the 
capture of neutron on Gadolinium releases a $8$ MeV gamma cascade. 
This being above the energy threshold, could be easy to observe in water 
detectors \cite{Beacom:2003nk}, transforming them 
into giant $\anue$ detectors with statistics many times 
the statistics expected in scintillator detectors. 
This could give exceptional sensitivity to neutrino 
oscillation parameters using reactor antineutrinos \cite{skgd}. 
This will help also in DSNB detection 
by lowering the lower energy threshold, 
and we should be able to use the same 
energy window as in liquid scintillators. Following 
\cite{Wurm:2007cy}, we present our results for the 
energy range $(10-30)$ MeV. 
The energy resolution of course continues to be given by 
Eq. (\ref{eq:eresolWC}).

\subsubsection{Liquid Argon Detectors}

Liquid argon TPCs are unique as they allow the 
detection of $\nue$. The only other 
$\nue$ sensitive detector technology 
that we have so far seen built on a large scale was the 
heavy water detector at SNO. 
However, SNO is now dismantled. Significant amount of R\&D on the 
other hand has gone into the liquid argon option. 
The ICARUS detector \cite{icarus} in 
Italy already consists of a $600$ ton module and has 
shown the feasibility of this detector technology. 
Since it is one of the few detector types 
which can be built on a large scale and allows for 
very fine granularity, good electron detection 
efficiency as well as 
detection of $\tau$ events, this is often considered as
a far detector option for the Neutrino Factory. 
Feasibility of probing galactic SN neutrinos was 
studied in  \cite{larsn,larsn_cs} and DSNB in 
\cite{Cocco:2004ac,Volpe:2007qx}. 
A future large liquid argon detector could have a mass of  
about $100$ kton. Some of the currently pursued 
proposals include GLACIER \cite{glacier}, MODULAr 
\cite{modular} and FLARE \cite{flare}.   
Energy resolution in this detector is expected to 
be extremely good and at energies relevant for 
the neutrino factory it is believed to be in the ballpark 
of $\sigma_E \sim 0.03\sqrt{E~{\rm (GeV)}}$
Therefore, at energies 
relevant for DSNB, one can assume 
that the energy reconstruction could be almost perfect. 
In what follows, we work under this assumption and give our 
results in terms of the neutrino energy. Since 
further R\&D would be needed to determine the 
backgrounds in this detector, we will show results for 
an energy window of $(20-40) $ MeV.

\section{Expected Events from DSNB}

\subsection{DSNB Antineutrino Events in Water and Scintillator Detectors}

\begin{table}
\begin{tabular}{|c|c|c|c|c|c|c|}
\hline
\hline
&&&&&&\\
& &SK&GDSK&HK&GDHK&LENA\\
Model&Hierarchy&(\footnotesize{19.3 - 30.0})
&(\footnotesize{10.0 - 30.0})&(\footnotesize{19.3 - 30.0})
&(\footnotesize{10.0 - 30.0})&(\footnotesize{10.0 - 25.0})\\
& & (MeV) & (MeV) & (MeV) & (MeV) & (MeV) \\
&&&&&&\\
\hline
\hline
\multirow{9}{*}{G1}& & & & & & \\
&NH&1.7 (1.7) &4.9 (4.9) &67.8 (67.8) &196.0 (196.0) &6.4 (6.4)\\
& & & & & &  \\
\cline{2-7}
& & & & & &  \\
&IH ($P_{13}=0$)&1.7 (2.7) &4.9 (7.4) &67.8 (109.6) &196.0 (296.0) &6.4 (9.5)\\
& & & & & &  \\
\cline{2-7}
& & & & & &  \\
&IH ($P_{13}=1$)&2.7 (1.7) &7.4 (4.9) &109.6 (67.8) &296.0 (196.0) &9.5 (6.4)\\
& & & & & &  \\
\hline\hline
\multirow{9}{*}{G2}& & & & & &  \\
&NH&1.1 (1.1) &3.5 (3.5) &42.6 (42.6) &139.5 (139.5) &4.6 (4.6)\\
& & & & & &  \\
\cline{2-7}
& & & & & &  \\
&IH ($P_{13}=0$)&1.1 (1.5) &3.5 (5.1) &42.6 (58.5) &139.5 (205.7) &4.6 (6.9)\\
& & & & & &  \\
\cline{2-7}
& & & & & &  \\
&IH ($P_{13}=1$)&1.5 (1.1) &5.1 (3.5) &58.5 (42.6) &205.7 (139.5) &6.9 (4.6)\\
& & & & & & \\
\hline\hline
\multirow{9}{*}{LL}& & & & & &  \\
&NH&2.5 (2.5) &6.2 (6.2) &98.2 (98.2) &246.0 (246.0) &7.7 (7.7)\\
& & & & & &  \\
\cline{2-7}
& & & & & &  \\
&IH ($P_{13}=0$)&2.5 (4.4) &6.2 (8.9) &98.2 (175.7) &246.0 (356.0) &7.7 
(10.6)\\
& & & & & &  \\
\cline{2-7}
& & & & & &  \\
&IH ($P_{13}=1$)&4.4 (2.5) &8.9 (6.2) &175.7 (98.2) &356.0 (246.0) 
&10.6 (7.7)\\
& & & & & &  \\
\hline\hline
\end{tabular}
\caption{\label{tab:events}
Number of expected events per year per 22.5 kton of SK and GDSK, 
1000 kton of HK and GDHK, and 50 kton of LENA. The events without 
collective effects are shown in parenthesis for comparison. 
}
\end{table}

\begin{figure}[t]
\includegraphics[height=1.0\textwidth,angle=270]{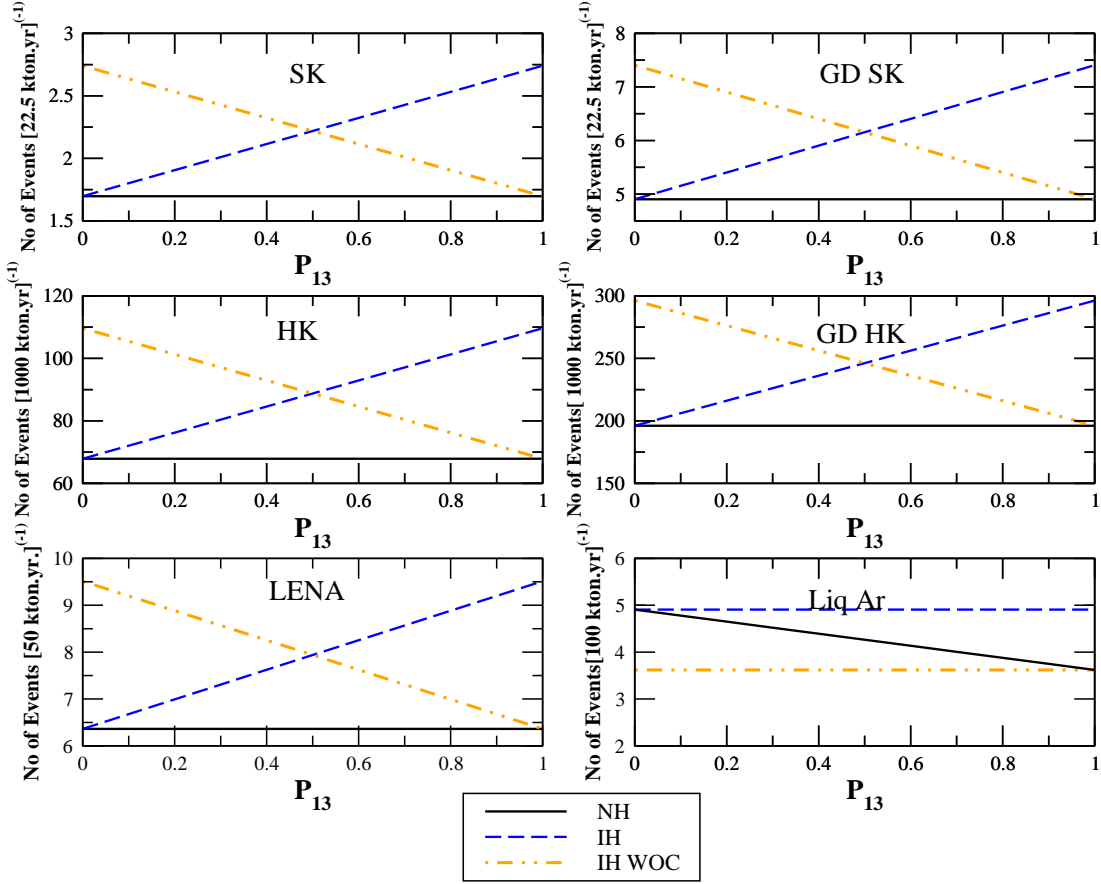}
\caption{\label{fig:eventsP13}
Number of expected events as a function of the jump 
probability $P_{13}$ for the G1 model. Black lines are for 
NH and blue dashed lines for IH. The yellow dashed dotted lines 
show the case for IH without collective effects
(WOC).}
\end{figure}

\begin{figure}[!h]
\includegraphics[height=1.0\textwidth,angle=270]{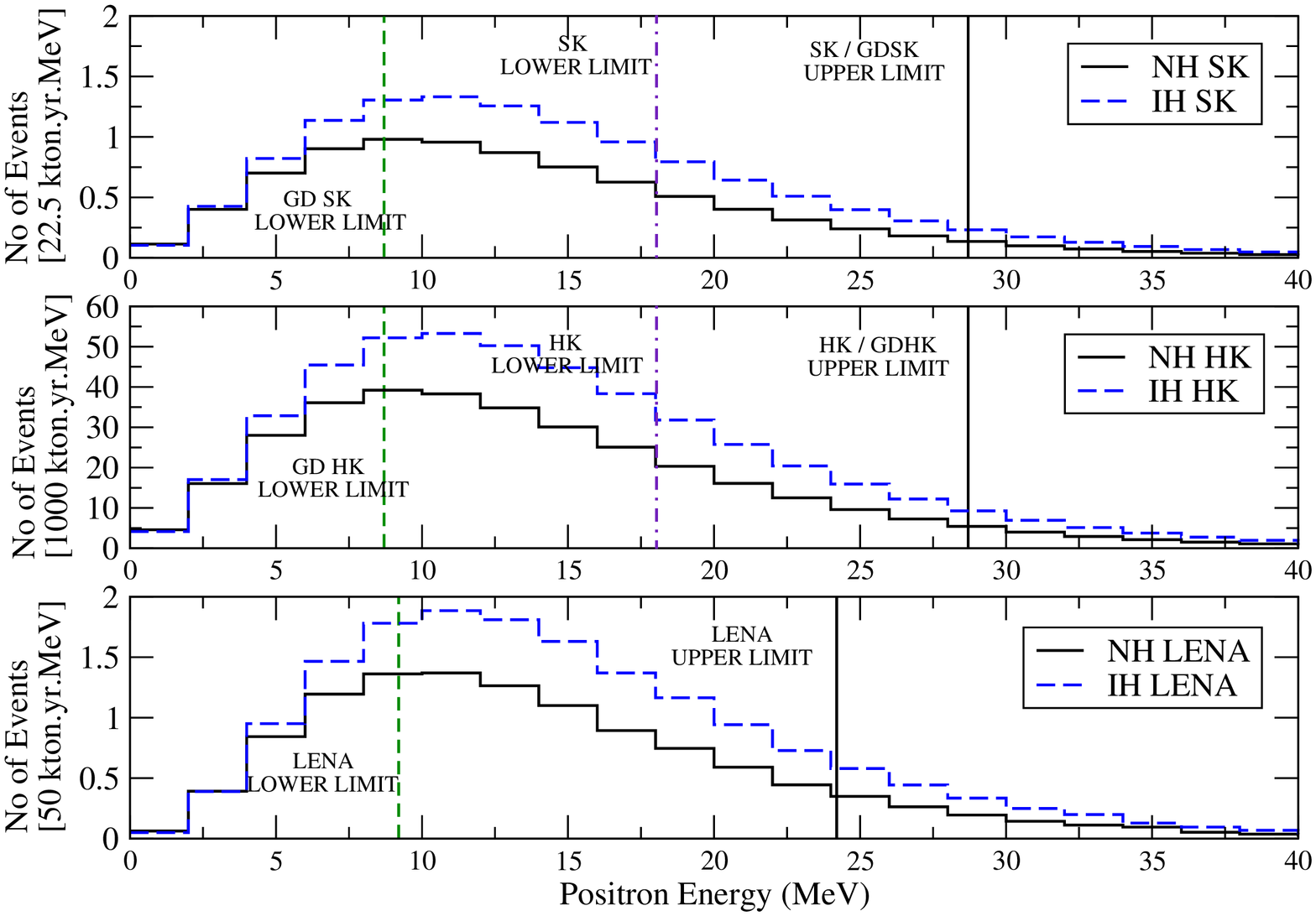}
\caption{\label{fig:eventsspec}
Number of expected events per year in 2 MeV 
positron energy bins
in SK (upper panel) 
HK (middle panel) and LENA (lower panel). The solid black lines 
show the projected event spectrum for NH while the dashed 
blue lines are if IH was true with $P_{13}=1$. The SN flux model 
corresponds to G1. 
}
\end{figure}

We give in this subsection the number of DSNB $\anue$ events expected in 
water and scintillator detectors. The total expected number of events 
are presented in Table \ref{tab:events}. The energy windows 
in which we have calculated the total number of events 
were discussed in the previous section and are 
shown in the parentheses in the first row of the table.
The number of events per year have been calculated assuming 
a fiducial mass of $22.5$ kton for SK and Gadolinium loaded SK (GDSK), 
$1$ Mton for a future megaton detector (marked in the 
table symbolically as HK) and Gadolnium loaded 
megaton water detector (GDHK) and $50$ kton for the scintillator detector (LENA). 
The results for NH remain the same for 
any value of $\theta_{13}$. For  
IH the neutrino oscillation probability and 
hence the number of events depend on $\theta_{13}$. 
We explicitly show results for two extreme values of 
$\theta_{13}$ --  
for small $\theta_{13}$ such that 
the jump probability $P_{13}=1$
and for large $\theta_{13}$ such that 
the jump probability $P_{13}=0$. 
For showcasing the impact of collective effects on 
the predictions for DSNB (anti)neutrino events, 
we also present in the Table \ref{tab:events} 
expected number of events if collective effects were not taken 
into account. These are shown in parenthesis. 
When there are no collective effects, one has only standard MSW 
transitions in the SN and it is well known that in 
this case antineutrinos 
undergo maximal flavor transitions for IH when 
$\theta_{13}$ is large ($P_{13}=0$), 
while for small values of $\theta_{13}$ 
or with NH (for any $\theta_{13}$) there is no 
matter enhanced resonant oscillations and 
these two scenarios give identical results. 
We therefore get larger number of events for 
IH and large $\theta_{13}$. 
However, 
once the collective effects are switched on, 
the small and large $\theta_{13}$ cases of IH 
switch roles. Since there are now two stages of 
flavor conversions, first due to collective effects deep 
inside the SN and then due to MSW transitions, 
the final $\anue$ fluxes are such that 
IH with large $\theta_{13}$ and NH 
give identical predictions, while IH with small $\theta_{13}$ 
predicts larger number of events 
(cf. upper right panel of Fig. \ref{fig:flux}). 

It can be seen, that we expect about a couple of 
events per year in SK \footnote{Note that there will also be a 
large number of background events in the detector and one has to 
find the signal by looking at excess of events 
above the fluctuations in the background. This makes DSNB detection 
more difficult.}.   
This would go up by a factor of about $2-3$ if Gadolinium were to be 
added to the water. The corresponding number for a megaton of 
water would be scaled upwards by a factor of $1000/22.5$ and we expect 
about $40-176$ ($140-356$) events per year in megaton water (Gadolinium 
loaded water) detectors depending on the choice of the neutrino 
mass hierarchy and $\theta_{13}$ and the SN model. After 
$10$ years of running these numbers would be a factor of $10$ higher, and 
we could have a few thousand events in the 
Gadolinium loaded detector. 
It should therefore be straightforward for megaton water 
detectors, with or without Gadolinium, to be able to observe 
these DSNB fluxes. More importantly we note that for a given 
SN flux model, it should be easy for megaton water detectors 
to determine the hierarchy, if $\sin^2\theta_{13}\ltap 10^{-5}$. 
For almost vanishing $\theta_{13}$, we can see that for G1, NH 
predicts $1960\pm 44$ ($678\pm 26$) 
events in 10 years of running of GDHK (HK) while 
IH predicts $2960 \pm 54$ ($1096\pm 33$). 
It would therefore be easy to distinguish 
one hierarchy from the other. Note that this is one of the very 
rare type of experiments which 
can give information about the neutrino 
mass hierarchy even if $\theta_{13}$ was below the reach of the 
most Neutrino Factory and Beta-beam experiments.   
A $50$ kton 
liquid scintillator detector should be able to record $46-106$
events in $10$ years of running. 

We have shown in the Table, number of events expected assuming 
either the G1, G2 or LL model for the initial SN 
neutrino fluxes. We find that the lowest number of events are predicted by 
the G2 model, while LL predicts the highest event rate. In fact, 
one can see that the event rate predicted by NH and G1 is 
close is that predicted by IH and G2. Likewise, the rate 
predicted by IH and G1 is close to the one predicted by 
NH and LL. 
We have discussed before the uncertainty associated 
with the SN models. Therefore, if the uncertainty in 
the model predictions for the initial fluxes remain at 
the current level, then it might be hard to distinguish the 
hierarchy from the DSNB itself, especially in the smaller detectors. 
However, for Gadolinium loaded 
megaton water detectors it might still be 
possible to say something about the hierarchy. 
Also, for G2 and NH (LL and IH) 
we have a prediction which is 
lower (higher) than any other case and therefore for these cases 
there is no confusion. For instance, if GDHK records 
less (greater) than $1500$ ($3000$) events, we could 
say that the hierarchy is normal (inverted). Of course, 
we have nowhere taken into account 
the uncertainty in the star formation rate. 
That might bring additional complication, which we do 
not address in this paper. 

So far we have presented results only for two extreme cases
of $\theta_{13}$, very low corresponding to $P_{13}=1$ and 
very high corresponding to $P_{13}=0$. For intermediate 
values of the mixing angle the jump probability ranges between 
0 and 1. We show in Fig. \ref{fig:eventsP13} how the total 
event rate in the different detectors change as a function of 
$P_{13}$. The SN model assumed is G1. 
Solid black lines show the case for NH while the 
dashed blue lines show the case for IH, where we have 
included both collective as well as MSW transitions inside the 
SN.  
It is easy to see from the expressions given in 
Table \ref{fluxtable} that for IH, 
the event rate would rise almost linearly with $P_{13}$.   
If collective effects were not taken into account then 
the trend would have been the opposite, and we would see a 
decrease in the $\anue$ event rate with $P_{13}$. These are shown 
for the different detectors by the yellow dot-dashed lines 
in the figure. 

For sizable number of events, it might even be possible to
do a spectral analysis of the DSNB events. We show in 
Fig. \ref{fig:eventsspec} the event spectrum for $22.5$ kton SK 
(upper panel), $1$ Mton HK (middle) and $50$ kton LENA (lower 
panel). The events per year are shown in $2$ MeV energy bins. The solid 
black lines give the event spectrum for NH while the 
blue dashed lines are for IH with $P_{13}=1$. We show results 
where both collective as well as MSW oscillations are taken 
into account. Upper and lower energy threshold for the different 
cases are indicated by vertical lines and we have assumed 
the G1 model for the initial fluxes.

\subsection{DSNB Neutrino Events in Liquid Argon Detectors}

\begin{table}
\begin{center}
\begin{tabular}{|c|c|c|c|}
\hline
\hline
&&&\\
& G1 & G2 & LL \\
&&&\\
\hline
\hline
&&&\\
NH ($P_{13}=0$) &  4.9 (4.9)   & 2.3  (2.3) & 9.9 (9.9)\\ 
&&&\\\hline
&&&\\
NH ($P_{13}=1$) & 3.6 (3.6)  &  1.7 (1.7) & 7.3 (7.3)\\
&&&\\\hline
&&& \\ 
IH  & 4.9 (3.6)  & 2.3 (1.7) & 9.9 (7.3)\\ 
&&&\\
\hline\hline
\end{tabular}
\end{center}
\caption{\label{tab:eventsnue}
Number of $\nue$ charged current 
events on $^{40}$Ar 
per year  per 100 kton of Liquid argon TPC in the 
energy window $E_\nu = (20-40)$ MeV. The events without 
collective effects are shown in parenthesis for comparison. 
}
\end{table}
\begin{figure}[!h]
\begin{center}
\vspace{-1.5cm}
\includegraphics[height=0.6\textwidth,angle=270]{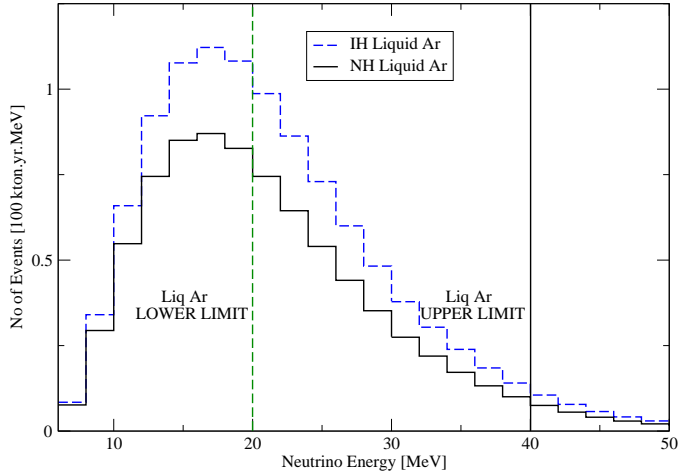}
\caption{\label{fig:eventslar}
Number of expected events per year in 2 MeV 
neutrino energy bins 
in a 100 kton Liquid argon TPC. The solid black lines 
show the projected event spectrum for NH ($P_{13}=1$) 
while the dashed 
blue lines are if IH was true. The SN flux model 
corresponds to G1. 
}
\end{center}
\end{figure}

Liquid argon TPC could offer a unique laboratory to probe 
$\nue$ from a future galactic SN as well as from the 
DSNB around us. We show in Table \ref{tab:eventsnue} 
the number of expected $\nue$ charged current events on 
$^{40}$Ar. We show results for NH and large $\theta_{13}$ 
($P_{13}=0$), NH and small $\theta_{13}$ 
($P_{13}=1$), and IH for any value of $\theta_{13}$. 
Expected number of events are shown for the 
three benchmark flux models. We see that the 
number of $\nue$ events expected in liquid argon detectors 
is extremely small. This is because the $\nue + ^{40}{\rm Ar} 
\rightarrow e^- + ^{40}{\rm K}^*$ cross-section (taken 
from \cite{larsn_cs}) is very small 
at low energies and rises very fast as the energy increases. 
The DSNB flux on the other hand gets redshifted to lower energies 
thereby reducing the number of events. In particular, the DSNB 
flux is peaked at around 5 MeV, with very few neutrinos 
in the energy window 20--40 MeV (cf. Fig. \ref{fig:flux}). 
It might also be interesting to compare the number of 
expected $\nue$ events 
in a liquid argon detector, with the number of 
$\anue$ events in a 
water detector, for same number of target nuclei/nucleons. 
It turns out that 100 kton of liquid argon has 
$1.5 \times 10^{33}$ argon targets, while 22.5 kton 
water detector (SK) also has $1.5 \times 10^{33}$ proton 
targets. On the other hand, 
the cumulative 
cross-section in the energy window of 20--40 MeV  
for $\nue$ capture on $^{40}$Ar 
is larger than the cross-section 
for $\anue$ capture on protons by a factor of about 2. 
Signal in this energy window, for the LL SN model  
with complete flavor 
conversion\footnote{For complete flavor conversion, the resultant 
flux at both liquid argon and water detector is $\nu_x$, and 
is therefore the same. Of course in reality, 
complete flavor conversion can be 
possible only in one channel. The above example is just to 
illustrate the difference in the number of events for the two 
detector types being compared.}, 
would be 9.9 and 5.8 events in 100 kton 
of liquid argon and 22.5 kton of water, respectively. 
This implies a ratio of about 1.7, which agrees with the rough 
estimate of the factor of 2 coming from the difference in the 
cross-sections.
If we could lower the energy threshold in liquid argon 
to 5.5 MeV, we could 
expect about 8.1 events per year 
for NH with small $\theta_{13}$ and 
about 10.5 events per year for IH (and NH with large $\theta_{13}$). 
In Fig. \ref{fig:eventslar}
we show the event spectrum in bins of 2 MeV width. The black solid line 
shows the spectrum for NH with $P_{13}=1$ while the blue dashed line 
is for IH.

\section{Summary and Conclusions}

Neutrinos emitted by core-collapse supernovae over the entire history of 
the Universe, pervade us. This is the so-called 
diffuse supernova neutrino background. 
The DSNB fluxes are theoretically given by folding the 
neutrinos emitted from a typical SN 
with the rate of SN explosions as a 
function of the redshift, and integrating over all redshifts to 
take into account all possible SN explosions that might have 
happened in the Universe. 
Since the final fluxes emerging from the SN depend on 
neutrino flavor coversions inside the SN, the 
DSNB fluxes also depend very crucially on neutrino properties. 
Therefore, while a galactic SN event is 
eagerly awaited in order to shed light on SN theory 
on one hand and neutrino properties on the other, detecting 
the DSNB in currently running and future detectors 
could help us constrain SN 
dynamics, cosmic star formation rate as well as neutrino properties. 
However, the primary agenda is to succesfully 
observe them in terrestrial detectors.

Being redshifted, the spectrum of 
DSNB fluxes is peaked at smaller energies, making their detection 
even more challenging. So far the running Super-Kamiokande 
detector has managed to put an upper bound on the $\anue$ 
DSNB flux. However, the situation might improve in the 
future with possibility of a signal in the upcoming large 
scale detectors which would be built to serve as 
the far detector for high performance neutrino beam 
experiments. Observing DSNB would be free for these detectors and 
the physics output from that would be immense. In this paper 
we re-analyzed the potential of a selected class of future 
detectors to detect DSNB fluxes. Such an analysis has been warranted 
by the flurry of activity in the field of SN neutrino research, 
following the revival of interest in neutrino-neutrino self-interaction. 
These interactions inside the SN have been shown to produce significant change to 
the final neutrino flux spectrum, especially if the neutrino mass hierarchy 
is inverted. Since these so-called collective effects 
inside the SN are unavoidable, 
it was necessary to revisit the issue of DSNB detection.

We considered water, Gadolinium loaded water and liquid 
scintillator detectors for $\anue$ DSNB detection and 
liquid argon TPC for observing the $\nue$ DSNB flux. A major 
issue in this field is the model uncertainties in the 
SN neutrino fluxes themselves.   
We presented results for three SN neutrino flux 
models. 
We calculated the total number of events for both the 
hierarchies and for two extreme values of $\theta_{13}$ 
resulting in jump probability $P_{13}\rightarrow0$ and 1. 
Number of events expected in future megaton water and 
50 kton liquid scintillator detectors are large, with a 
few thousand events expected in Gadolinium loaded 
megaton water detectors running for 10 years. 
For true inverted hierarchy, it becomes possible to 
get very large flavor oscillations even if $\theta_{13}\rightarrow0$. 
We showed that under fortunate circumstances, it might 
be possible to get information on the neutrino mass hierarchy
by observing DSNB in megaton water detectors. Note that 
this is a very unique situation, since for $\theta_{13}\rightarrow0$ it 
becomes almost impossible to determine the hierarchy using long baseline 
experiments. In this way, DSNB detection could be complementary to 
the long baseline program. We also showed how the 
total number of events change if $\theta_{13}$ increases from 
very small to very large values, decreasing $P_{13}$ from $1$ to $0$. 
Finally, we showed the event spectrum by binning the prospective 
data in $2$ MeV bins. 

In conclusion, very large number of DSNB events are expected 
in the next generation detectors and therefore, 
it should be possible to observe DSNB 
$\anue$ in the future. 
Collective effects inside 
SN significantly change the predicted number of DSNB events if 
the hierarchy is inverted. Under fortunate conditions 
it might be possible to determine the neutrino mass hierarchy 
using ths DSNB signal and this could be done 
even if $\theta_{13}\rightarrow0$, 
in which case long baseline experiments would not be able to 
tell the hierarchy at all. Propects of DSNB detection look 
extremely promising and one might even feel optimistic about 
learning about neutrino oscillation parameters, cosmic star formation 
rate and maybe about SN physics, by observing these 
relic neutrinos in future detectors.


\vglue 0.8cm
\noindent
{\Large{\bf Acknowledgments}}\vglue 0.3cm
\noindent
Authors wish to thank Manoj Kaplinghat for discussions 
during very early stages of this work, and 
the organizers of WHEPP X, during which this work was initiated.   
This work has been supported by the Neutrino Project
under the XI Plan of Harish-Chandra Research Institute and 
the project 'Frontiers of Theoretical Physics' under the 
XI Plan of Saha Institute of Nuclear Physics. 
Basudeb Dasgupta acknowledges partial support 
from the Max Planck India Partnergroup at Tata 
Institute of Fundamental Research. Sovan 
Chakraborty would like to thank 
Harish-Chandra Research Institute for kind and warm 
hospitality. The authors thank Sebastian Galais and Christina Volpe
for their critique of the previous version of this paper.


\end{document}